\pdfoutput=1
\documentclass[twocolumn,secnumarabic,amssymb, nobibnotes, aps, prl, superscriptaddress]{revtex4-2}

\newcommand{\ydr}[1]{{\color{black} #1}}
\newcommand{\rev}[1]{{\color{black} #1}}

\usepackage{graphicx}
\usepackage{dcolumn}
\usepackage{bm}
\usepackage{amsmath}
\usepackage{enumerate}
\usepackage{setspace}
\usepackage{dsfont}
\usepackage{subfigure}
\usepackage{multirow}
\usepackage{indentfirst} 
\usepackage{mathrsfs}
\usepackage{color}
\usepackage{lineno}
\usepackage{threeparttable}  
\usepackage{cuted}
\usepackage[pdfstartview=FitH,
CJKbookmarks=true,
bookmarksnumbered=true,
bookmarksopen=true,
colorlinks, 
pdfborder=001,
linkcolor=blue,
anchorcolor=blue,
citecolor=blue
]{hyperref}

\begin{document}

\title{Time-interval Measurement with Linear Optical Sampling at the Femtosecond Level}%

\author{Dongrui Yu}%
\affiliation{State Key Laboratory of Advanced Optical Communication Systems and Networks, School of Electronics, and Center for Quantum Information Technology, Peking University, Beijing 100871, China}

\author{Ziyang Chen}%
\email[Email: ]{chenziyang@pku.edu.cn}
\affiliation{State Key Laboratory of Advanced Optical Communication Systems and Networks, School of Electronics, and Center for Quantum Information Technology, Peking University, Beijing 100871, China}

\author{Xuan Yang}%
\affiliation{State Key Laboratory of Information Photonics and Optical Communications, Beijing University of Posts and Telecommunications, Beijing 100876, China}

\author{Yunlong Xu}%
\affiliation{State Key Laboratory of Information Photonics and Optical Communications, Beijing University of Posts and Telecommunications, Beijing 100876, China}

\author{Ziyi Jin}%
\affiliation{State Key Laboratory of Information Photonics and Optical Communications, Beijing University of Posts and Telecommunications, Beijing 100876, China}

\author{Panxue Ma}%
\affiliation{State Key Laboratory of Information Photonics and Optical Communications, Beijing University of Posts and Telecommunications, Beijing 100876, China}

\author{Yufei Zhang}%
\affiliation{State Key Laboratory of Advanced Optical Communication Systems and Networks, School of Electronics, and Center for Quantum Information Technology, Peking University, Beijing 100871, China}

\author{Song Yu}%
\affiliation{State Key Laboratory of Information Photonics and Optical Communications, Beijing University of Posts and Telecommunications, Beijing 100876, China}

\author{Bin Luo}%
\email[E-mail: ]{luobin@bupt.edu.cn}
\affiliation{State Key Laboratory of Information Photonics and Optical Communications, Beijing University of Posts and Telecommunications, Beijing 100876, China}

\author{Hong Guo}%
\email[E-mail: ]{hongguo@pku.edu.cn}
\affiliation{State Key Laboratory of Advanced Optical Communication Systems and Networks, School of Electronics, and Center for Quantum Information Technology, Peking University, Beijing 100871, China}

\date{\today}%

\begin{abstract}
	High-precision time-interval measurement is a fundamental technique in many advanced applications, including time and distance metrology, particle physics, and ultra-precision machining. However, many of these applications are confined by the imprecise time-interval measurement of electrical signals, restricting the performance of the ultimate system to a few picoseconds, which limits ultra-high-precision applications. Here, we demonstrate an optical means of the time-interval measurement of electrical signals that can successfully achieve femtosecond (fs)-level precision. The setup is established using the optical-frequency-comb (OFC)-based linear optical sampling technique to realize timescale-stretched measurement. We achieve the measurement precision of 82 fs for a \rev{single LOS scan} measurement and 3.05 fs for the 100-times average with post-processing, which is three orders of magnitude higher than the results of older electrical methods. The high-precision time interval measurement of electrical signals can substantially improve precision measurement technologies.
\end{abstract}

\maketitle


\section{Introduction}

\begin{figure*}[ht]
	\centering
	\includegraphics[width=\linewidth]{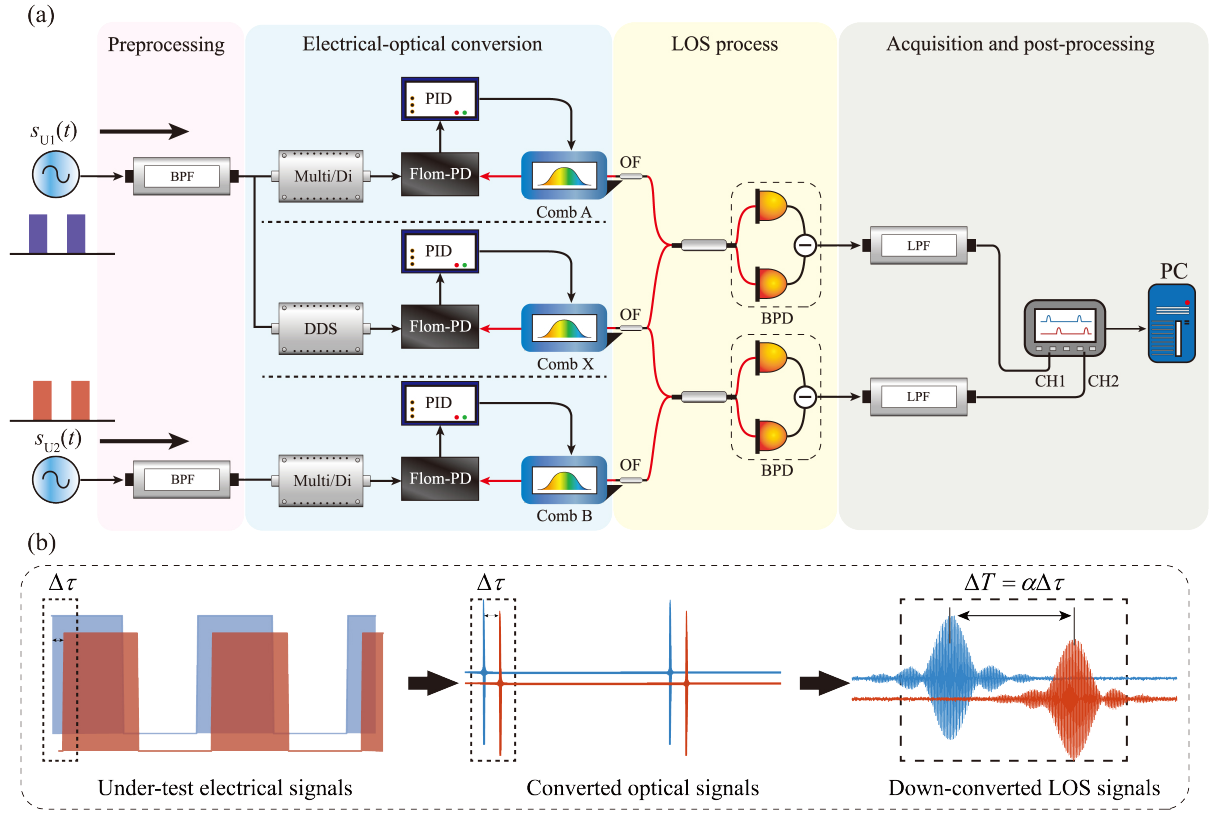}
	\caption{Schematic of the time-interval measurement setup. (a) The time-interval measurement setup. The two band-pass filters (BPF) in the preprocessing part filter \ydr{a certain harmonic} of the under-test pulses, which serve as the combs' reference after being multiplied, divided, or synthesized. The direct digital synthesizer (DDS) in \ydr{the electrical-to-optical} conversion part generates a sinusoidal wave with a slight frequency difference from its input signal. The repetition \ydr{frequencies} of three combs are phase-locked with the fiber-loop optical-microwave phase detector (FLOM-PD) and the PID. The outputs of comb A and B are sampled through LOS with comb X, and then detected and acquired by balanced photodetectors (BPDs) and an oscilloscope. The low-pass filters (LPF) are used to \ydr{isolate} the final output. (b) Conversion of the under-test electrical pulses to the measured LOS signals. Through preprocessing and electrical-optical conversion, the signals are converted to two optical-frequency-comb signals, carrying the same time-interval information \ydr{as} the under-test signals. Through the LOS process, the optical signals are down-converted to two LOS signals in terms of the frequency, whose time intervals are amplified $\alpha$ times in comparison with the under-test signals.}
	\label{system}
\end{figure*}

High-precision time measurement has significant applications in fields including remote time synchronization~\cite{timesync1,timesync2,timesync3,timesync4} and precise navigation systems~\cite{navig1,navig2,navig3,navig4}.
A prominent example is the research on numerous big-scientific experiments, where high-precision time measurement can directly limit the experimental performance, such as time-of-flight measurement in high-energy particle physics~\cite{highenergy1}, dark matter detection and measurement of fundamental constants~\cite{darkmatter1,darkmatter2,darkmatter3,darkmatter4,const1,const3,const4}.
Specifically, high precision is always the ultimate goal of time measurement, and the progression of its technique can help overcome technical bottlenecks in precision measurement in relevant fields and can also bring about a breakthrough in discovering new areas of physics, such as particle physics and cosmology.

Traditional solutions for measuring the time interval of electrical signals use electrical means~\cite{RN2390,RN2377,RN2381}; they measure the equivalent voltage proportional to the under-test time interval~\cite{highenergy1,RN2392} or use the digital counting method to directly count the number of a clock's cycle~\cite{highenergy1,RN2392,RN2386,RN2391}. The precision of the former method, with the order of hundred picoseconds (ps), is limited by the poor resolution of digital devices; the latter methods, with the help of the dual mixer time difference measurement technique, can achieve better performance than 0.1 ps~\cite{RN1156}. However, such a sub-ps-level time-interval measurement cannot satisfy the requirements of state-of-the-art precise implementations, such as gravitational wave detection and femtosecond (fs)-level time-frequency transfer.
To support the rapidly growing precision requirement and the expanding directions of applications, researchers must urgently find a breakthrough approach for precision measurement.

Compared with traditional electrical techniques, optical-frequency-comb (OFC)-based techniques offer a possibility for high-performance time-frequency measurement owing to their ultra-high time resolution~\cite{OFC1,OFC2,RN2266,RN2265}; therefore, this technology enables high-precision time-interval measurement.
Remarkably, dual-OFC-based linear optical sampling (LOS) techniques have been extensively studied in many high-precision precise measurement fields. For example, fs-level remote time synchronization~\cite{timesync1,LOSsync1,LOSsync2} and nanometer absolute distance measurement~\cite{dist1,dist2,dist3,dist4} have been achieved, showing the advantage of using optical means for precision measurement.
However, the time-interval measurement of electrical signals by optical methods faces numerous technical difficulties, such as the noise introduced by electro-optical conversion, the precision of optical methods, the robustness and complexity of the optical system, and the capability of measuring irregular waveforms, which leads to challenges for high-performance and simple system design.

Here, we demonstrate an optical method for ultra-high-precision time-interval measurement that measures two \rev{periodic} electrical signals with fs-level precision, despite arbitrary waveforms. The under-test electrical pulses were locked to the repetition frequencies of two OFCs to transfer the interval information from the electrical to the optical region. By introducing dual-comb LOS technology, we used a third OFC locked to one of the under-test signals to sample the two optical pulses, which in principle overcomes the limitation of electrical means. As the carrier-envelope offset frequency does not need to be manually stabilized, the complexity of the entire system was significantly reduced. We showed that the precision of the time-interval measurement of electrical pulses fell below 3.05 fs after LOS and post-processing. The findings demonstrate the feasibility of the fs-level time-interval measurement of electrical pulses and provide support for ultra-high-precision microwave applications.


\section{Results}
\label{sec:results}

\textbf{Experimental setup.}
\label{sec. experimentalsetup}
A schematic diagram of the time-interval measurement system is illustrated in FIG.~\ref{system}~(a). When measuring two input electrical periodic signals with arbitrary waveforms, traditional electrical methods are difficult to quantify with sufficient precision. For this reason, our proposed optical protocol performs four procedures \ydr{to achieve better measurement precision}: signal preprocessing, conversion from electrical to optical signals (also called E-O conversion for simplicity), LOS processing, and data acquisition and post-processing.

\rev{In the signal preprocessing phase, the system uses adjustable band-pass filters (BPFs) and frequency multipliers/dividers to obtain a single harmonic component of the input waveforms that equals integer multiples of 100 MHz, for the subsequent process.
Note that the phase time of every harmonic contains the same information as the original electrical signal dose~\cite{RN928} (see Methods section).}
Measuring \ydr{the} time interval of the sine signal directly by electrical means leads to the precision of ps level~\cite{RN964, RN2539}. To overcome this limitation, in the E-O conversion process, we lock the repetition frequencies of two OFCs, namely comb A and comb B, to the \ydr{previously obtained single harmonic} signals using the fiber-loop optical-microwave phase detector (FLOM-PD) technique~\cite{FLOM-PD1, RN892, RN1150}. 
\ydr{In our scheme, the 2 GHz} harmonic of the under-test signal is \ydr{obtained} and used for phase locking, and the residual phase noise after phase locking was \ydr{$8\times10^{-14}$} @ 1 s, as evaluated by the phase noise analyzer (\textit{MicroChip 53100A}). It ensures a standard deviation of less than 45 fs of the locking-induced noise of the time measurement. 

In the optical domain, the system uses the state-of-the-art LOS technique to down-convert two comb signals to microwave combs with $\alpha$ times precision improvement (see the Methods section). The key parameter $\alpha= {{{f_{\rm{r}}}} \mathord{\left/{\vphantom {{{f_{\rm{r}}}} {\Delta {f_{\rm{r}}}}}} \right.\kern-\nulldelimiterspace} {\Delta {f_{\rm{r}}}}}$ quantifies the stretch of time scale when using a local oscillator (LO) comb (comb X in FIG.~\ref{system}~(a)), with a slightly different frequency of ${\Delta {f_{\rm{r}}}}$, to sample the measured OFC with the frequency ${f_{\rm{r}}}$, yielding a microwave comb containing the time information of the measured OFC, but with higher precision. The frequency difference is achieved by a direct digital synthesizer (DDS).
Here, we used two OFCs with \ydr{a} repetition frequency of 100 MHz as the measured OFCs (comb A and B) and an OFC with \ydr{a} repetition frequency of 100.001 MHz as the LO-OFC, which improved the precision of time measurement by $\alpha  = {10^5}$ times.
\rev{It is worth noting that the repetition frequencies of the combs are all adjustable with a temperature controller and a PZT, enabling them to be synchronized to the under-test signals for the subsequent process.}

In the LOS process, the optical part is constructed with full-polarization-maintaining fiber couplers and optical filters (OFs) with a central wavelength of 1550.12 nm and a bandwidth of 0.8 nm to prevent polarization-induced amplitude fluctuation.
OFs are used not only to limit the OFCs' spectrum in the same band to obtain a stable and clear interference pattern but also to fulfill the Nyquist condition and to prevent spectrum overlapping~\cite{RN2333}.
The interference result is subsequently transformed into an electrical signal with balanced photodetectors (BPDs, \textit{Thorlabs PDB450C}) for the convenience of data acquisition and processing. Therefore, the conversion process can be summarized by first converting under-test electrical signals to optical signals, then down-converting to the microwave LOS signals using the LOS technique, as illustrated in FIG.~\ref{system}~(b).
\rev{In the whole process, to ensure the long-term stability of the system, we shortened the pigtails of commercial WDM devices, couplers, and other optical components and directly fused them together. We also kept the electrical cables as short as possible and implemented temperature control in the detection section to minimize the impact of internal path variations on time measurements. The upper and lower branches of the system have a baseline delay for not precisely matching the path length, while it is calibrated before conducting the measurements.}

\begin{figure}[ht]
	\centering
	\includegraphics[width=0.9\linewidth]{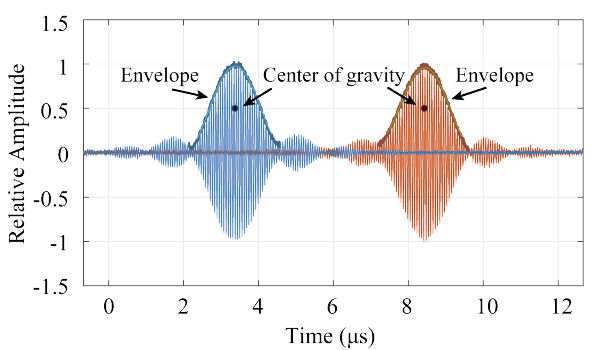}
	\caption{\ydr{Processing} of the down-converted LOS signal by the envelope calculation and \rev{localization (i.e., center finding)}. The process is achieved by first fitting \ydr{the} signal's envelope and then calculating \ydr{its} center of gravity.}
	\label{signalconverting}
\end{figure}

The acquisition part contains two electrical low-pass filters (81 MHz) to obtain the frequency components introduced by the beating OFCs, and an oscilloscope (OSC, \textit{Rigol MSO8204}) is used to simultaneously acquire the two channels' data. 
The duration of an LOS signal in our experiment was approximately 4 $\mu$s. To fit the envelope of the LOS signal accurately, we expected at least 100 points in the pulse duration. This means that the sampling rate was expected to be over $100\ \text{pts}/4\ \mu\text{s} = 25\ \text{MHz}$.
A higher sampling rate will improve the resolution of the LOS signal (see the Methods section), while \ydr{it will also} increase the cost of data processing. To balance the LOS signal precision and the processing speed, we set the sampling rate to 500 MHz in our experiment.
A screen containing \rev{a set of the LOS waveforms} is shown in FIG.~\ref{signalconverting}, where two input channels of the OSC are simultaneously acquired and then delivered to the computer. These two signals \ydr{contain} time information of the input electrical signals, but with higher measurement precision.
The timescale of the down-converted LOS signal is at the microsecond level, which can be easily acquired and processed using digital signal processing techniques, compared with the timescale of tens of picosecond level with \ydr{direct} measurement. The data processing procedure is given in the Methods section.

\begin{figure*}[ht]
	\centering
	\includegraphics[width=0.9\linewidth]{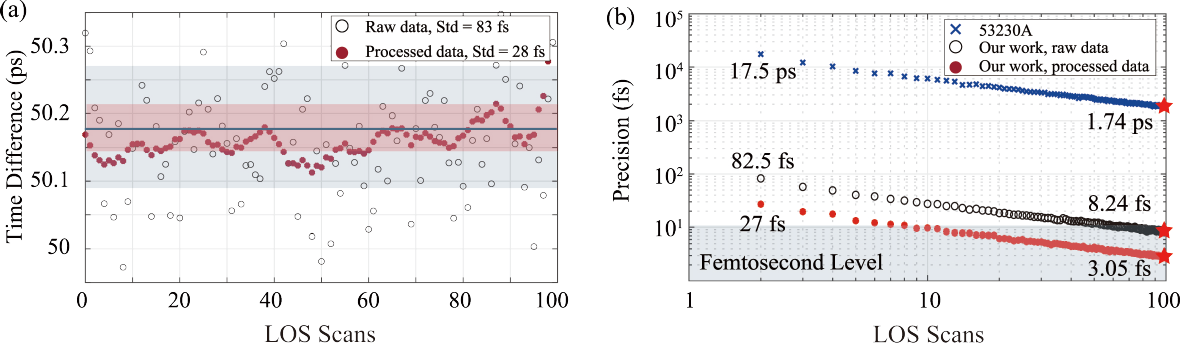}
	\caption{Measurement results of system's precision. (a) Measurement results of the time difference. Black circles represent raw data, while red points represent data processed with the Kalman filter. (b) Measurement precision of the data. Blue crosses \ydr{indicate} the precision of a commercial time-interval counter (Keysight, 53230A). Black circles represent the precision of our work with raw data, while red points represent the precision of our work with processed data.}
	\label{precision}
\end{figure*}

\begin{figure}[ht]
	\centering
	\includegraphics[width=0.9\linewidth]{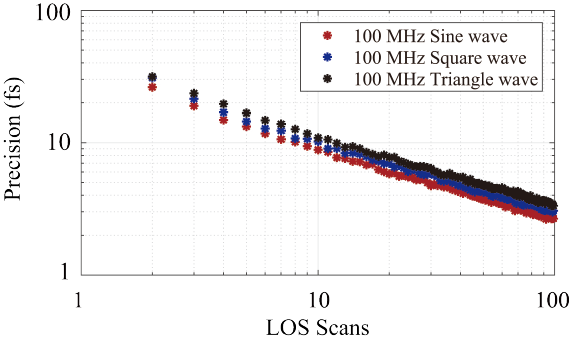}
	\caption{The measurement precision (processed data) of the system using 100 MHz sine waves, square waves, and triangular waves. The result shows similar measurement precision for the different waveforms.}
	\label{arbitrarywaveform}
\end{figure}

\textbf{Measurement precision.}
In our experiment, we assessed the precision of the time interval measurement system by generating the test signal through power-splitting a single 100 MHz square wave produced by an arbitrary waveform generator (AWG, \textit{Keysight M8195A}). 
As explained in Sec.~\ref{sec:results}~\ref{sec. experimentalsetup}, we applied filtering to isolate the first harmonic of the signals, which were then multiplied to 2 GHz. These resulting signals served as the references for comb A and comb B, respectively. 
Hence, the under-test time interval of two square waves was transformed into the measurement of two down-converted LOS signals. In our scheme, we measured the interval by calculating the LOS envelopes’ center of gravity, as shown in FIG.~\ref{signalconverting}, and the calculated result is shown in FIG.~\ref{precision}~(a). With continuous acquisition, the time interval could be displayed and plotted in \ydr{real-time}. The black circles stand for the raw measured data, with the average value being -50.15 ps and the standard deviation being 83 fs. To further improve the precision, we also optimized the system's performance with a Kalman filter in \ydr{real-time}. The red dots stand for the processed data with the Kalman filter, with the standard deviation being 28 fs.

\begin{figure*}[t]
	\centering
	\includegraphics[width=\linewidth]{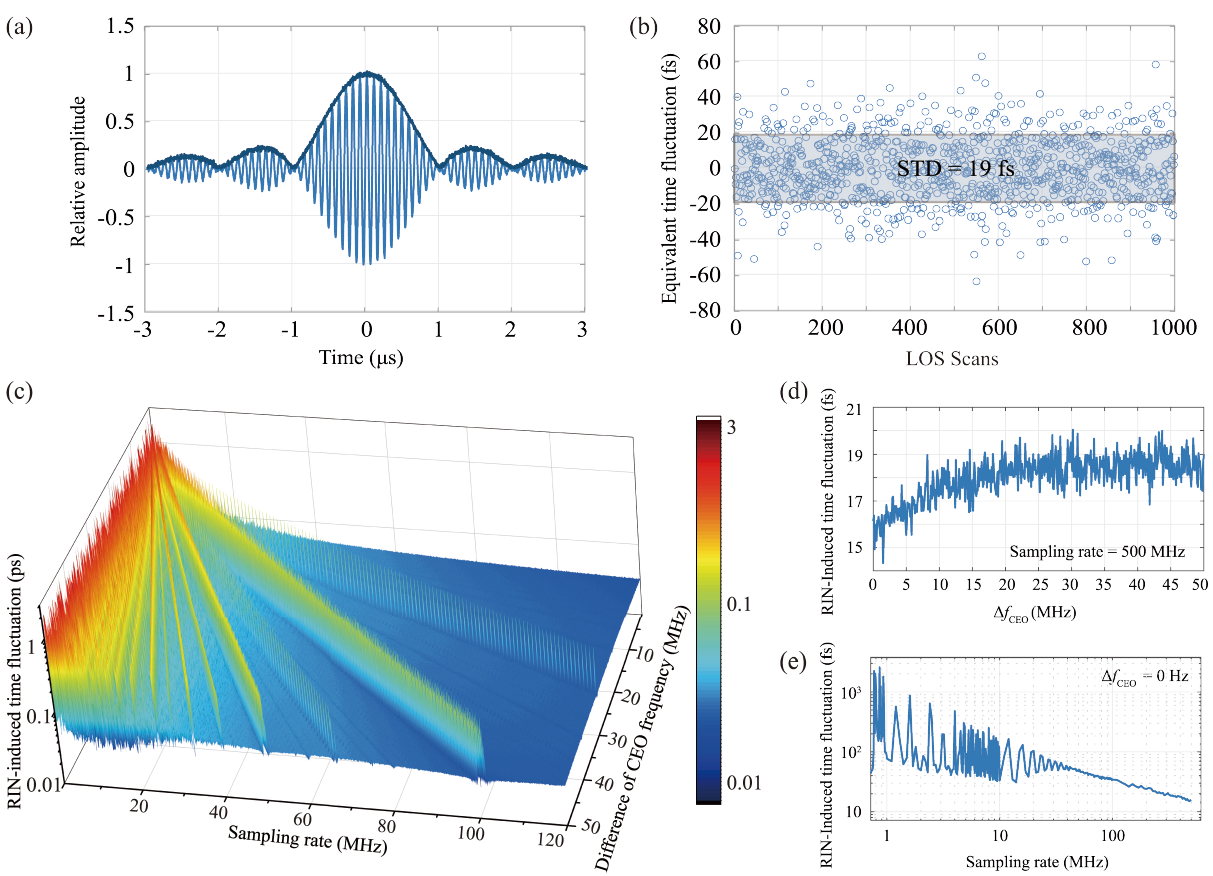}
	\caption{Simulation results of the effect of the relative intensity noise (RIN)-induced time fluctuation. (a) The RIN affects the \rev{localization} process and brings fitting \ydr{errors}. (b) The time fluctuation \ydr{is} caused by 1.32\% RIN, where the sampling rate is set to 500 MHz corresponding to \ydr{the} experimental parameter. (c) The relation of RIN-induced time fluctuation, $f_\text{CEO}$-difference, and sampling rate for a fixed RIN level 1.32\%. (d) The RIN-induced time fluctuation under the sampling rate of 500 MHz. (e) The RIN-induced time fluctuation under the $f_\text{CEO}$-difference of 0 Hz.}
	\label{3d}
\end{figure*}

To further reduce the system noise, we also considered averaging the sampling \ydr{results} to obtain the optimized results. Here, we used the standard error of the mean (SEM) to estimate the precision after averaging, given by
\begin{equation}
	\text{SEM}=\sigma(T)/\sqrt{n},
\end{equation}
where $\sigma$ represents the standard deviation of the data, and $n$ is the number of \rev{scans of the LOS measurement}. We estimated the system's performance after averaging, and the precision is shown in FIG.~\ref{precision}~(b).
As the \rev{scanning} number increased, the sample mean gradually approached the expectation value with a slope of tangent of -1/2, and the measurement precision reached 8.24 fs with \ydr{an} average of 100 raw data and 3.05 fs with \ydr{an} average of 100 Kalman-filtered data.

\rev{Furthermore, our scheme is capable of measuring the time interval of periodic signals with arbitrary waveforms. To demonstrate this, we generated sine waves, square waves, and triangular waves using the AWG. The consistent measurement results presented in FIG.~\ref{arbitrarywaveform} provide evidence to support this claim.}

\ydr{The} precision was achieved via 100-times acquisition, data transmission, and data processing, and it took approximately 1 second per cycle, with time being mainly consumed in the transmission part. When the OSC was replaced with a high-speed acquisition card, the system achieved 1,000 \rev{LOS scans} of output per second, and the fs-level precision was realized in 0.1 \ydr{seconds}.

\textbf{Key parameters analysis.}
To investigate the noises introduced by different components, we analyzed the key elements influencing the performance of our system. 
The system's performance was limited by various factors, including the synchronization precision of combs A and B from the under-test signal, the amplitude noise from the detectors, and the sampling rate.

The amplitude noise and sampling rate affected the system precision during the \rev{localization} process. To investigate their influence in detail, we simulated them with the LOS model described in the Methods section.
In our experiment, the BPD and the light source introduced amplitude noise with \ydr{a} standard deviation of 7 mV, which is approximately 1.32\% of an LOS pulse, and we denote it as the relative intensity noise (RIN). The RIN affects the \rev{localization} process and thus brings fitting error to the measurement result, as shown in FIG.~\ref{3d}~(a). We used 1.32\% Gaussian white noise added on an LOS signal's amplitude to simulate the system's performance. The sampling rate was set to 500 MHz, as determined by experimental parameters, and the standard deviation of the calculated time-interval fluctuation was 19 fs (see FIG.~\ref{3d}~(b)).

Note that the influence of RIN noise was tightly connected with both the $f_\text{CEO}$ (carrier-envelope offset frequency, see Methods section) difference between the two combs of an LOS signal and the sampling rate of the OSC. Because $f_\text{CEO}$ of the combs are all \ydr{free-running}, it is crucial to choose an appropriate sampling rate to reduce the influences brought by the drift of $f_\text{CEO}$ and also take the processing speed into account.
For that purpose, we simulated the relation of RIN-induced time fluctuation, $f_\text{CEO}$-difference, and sampling rate for a fixed RIN level 1.32\%, and the results are shown in FIG.~\ref{3d}~(c). The y-axis ranges from 0 to 50 MHz because of the 100 MHz repetition frequency of the combs; a larger $\Delta f_\text{CEO} = f_\text{r}/2+\delta f$ is practically equivalent to $\Delta f_\text{CEO} = f_\text{r}/2-\delta f$ owning to the periodicity of the combs' spectrum.

The results of FIG.~\ref{3d}~(c) illustrate that when the sampling rate was larger than the repetition frequency, namely, 100 MHz, the RIN-induced time fluctuation was not evidently related to $\Delta f_\text{CEO}$. The \rev{higher} the sampling rate is, the lower \ydr{the} noise would remain. The 500-MHz sampling rate in our experiment led to 16--20 fs noise, as shown in FIG.~\ref{3d}~(d), which is relatively low compared with the phase-locking-induced noise.
If the sampling rate was below 100 MHz, the time-interval fluctuation level would be affected owing to the varying $f_\text{CEO}$ of the combs. When $\Delta f_\text{CEO}$ was around a half-integer multiple of the sampling rate, the noise significantly increased because of the inaccurate sampling of the pulse. In addition, when the sampling rate fell lower, the noise level of other points also gradually increased because the amplitude fluctuation cannot be effectively averaged out with so few points.
\rev{Moreover, to further improve the localization precision, 
increasing the optical bandwidth of the OF and equivalently decreasing the pulse width could potentially make a significant contribution.}

According to the system's model given in the Method section, the measurement noise floor was mainly composed of the residual phase noise introduced by phase locking, as well as the RIN-induced time fluctuation. That is,
\begin{equation}
	{\sigma _{{\rm{floor}}}} = \sqrt {\sigma _{{\rm{P}}{{\rm{L}}_{\rm{A}}}}^2 + \sigma _{{\rm{P}}{{\rm{L}}_{\rm{B}}}}^2 + \sigma _{{\rm{RI}}{{\rm{N}}_{\rm{A}}}}^2 + \sigma _{{\rm{RI}}{{\rm{N}}_{\rm{B}}}}^2} ,
\end{equation}
where $\sigma$ denotes the standard deviation of residual phase noise. The locking-induced noise was measured by ${\sigma _{{\rm{P}}{{\rm{L}}_{\rm{A}}}}} = {\sigma _{{\rm{P}}{{\rm{L}}_{\rm{B}}}}} = 45$ fs, and the RIN-induced noise was measured by ${\sigma _{{\rm{RI}}{{\rm{N}}_{\rm{A}}}}} = {\sigma _{{\rm{RI}}{{\rm{N}}_{\rm{B}}}}} = 19$ fs, which yielded the noise floor of the system of 69 fs. This calculated result indicates that the dominant noise in the existing system is locking-induced noise, which can be further optimized using a higher-order harmonic for the FLOM-PD's phase tracking. The \ydr{experimentally} measured noise floor of the system was 82 fs, which is consistent with the calculated result. 
The difference between the calculated noise floor and the measured noise floor may have stemmed from the fluctuation of the amplification parameter $\alpha$ caused by the small fluctuation of the DDS-generated frequency difference $\Delta f_\text{r}$.

Interestingly, in FIG.~\ref{3d}~(c), for some fixed $\Delta f_\text{CEO}$, there were still low-sampling-rate regions with relatively low RIN-induced time fluctuation. For instance, if $\Delta f_\text{CEO}$ could be manually controlled to 0 Hz, the regions that have large noise (the green regions) could be avoided, even with low sampling rates, as shown in FIG.~\ref{3d}~(e). \ydr{In this situation}, the sampling rate of 100 MHz was sufficient to achieve RIN-induced noise below 40 fs (below the phase-locking-induced noise), although the system complexity would \ydr{worsen}.
\rev{It is noteworthy that the oscillations observed in FIG.~\ref{3d}~(d)(e) are likely a result of the manually added Gaussian noise. Therefore, our focus lies on the overall magnitude rather than the specific values of individual data points.}

\rev{For future studies, since the RIN-induced localization noise is at a relatively low level and has limited influence on the system performance at this stage, we did not conduct further in-depth research in this aspect and experimentally treated it as a known factor (the experimentally measured value is 1.32\%). However, it is a complex composite at least consisting of analog-digital-conversion quantization accuracy, detector electrical noise, and shot noise. It will be an important theoretical question to be analyzed and calculated after improving the accuracy of light source locking.}

\rev{In other aspects, the update rate (equal to $\Delta f_\text{r}$) is also a crucial consideration in some applications. It can be increased by simply enlarging $\Delta f_\text{r}$. However, it cannot be infinitely large considering the LOS amplification factor $\alpha$ and the Nyquist bandwidth. Exploring the trade-off between these factors would be a valuable area for further research.}

\textbf{Discussion.}
In this study, with the optical mean, namely, the LOS technique, we
achieved a high-precision and simple time-interval-measurement setup that can drastically
promote the precise measurement of electrical signals. For demonstration, we built a CEO-free system for measuring the time interval of two squared electrical signals, and simulations were implemented to optimize the experimental parameters. Experimentally, we achieved 83 fs in a single-LOS-scan precision, and we further approached 3.05 fs with data processing, which is about three orders of magnitude better than the result of traditional time-interval counter methods.

Compared with traditional direct detection of the time interval, our method has the advantage of avoiding the ambiguity brought by the rising edge of a pulse, creating the potential for the time-interval measurement of \ydr{arbitrary waveform pulsed signals}. The possible restrictions of the performance were also discussed theoretically, and they can guide us toward further optimization of the setup.
\rev{In our approach, the measurement is performed by synchronizing the combs with the periodic signals under test. As a result, we are currently unable to measure the time interval between individual pulses, and this aspect is left for future investigation.}
Based on our work, a real-time high-precision time-interval measurement system can be developed to achieve sub-fs measurement, which we believe will open a new pathway for high-precision metrology, such as ultra-high-precision navigation and time synchronization.


\section{Methods}

\textbf{Basic principles of the LOS technique.}
The LOS technique is achieved by the interference between three OFCs, whose spectrum comprises a series of narrow-bandwidth lines with the frequency $f_n$, which can be written as $f_n = f_\text{CEO} + n f_\text{r}$, where $n$ represents large integers (typically with the order of $10^6$, with $f_n$ being in an optical region). Moreover, $f_\text{r}$ denotes the repetition frequency, and $f_\text{CEO}$ is the carrier-envelope-offset (CEO) frequency, which is related to the carrier and envelope phase offset. 
In the time domain, a pulse train with a short duration and a period of $1/f_\text{r}$ is generated by an OFC, and the ultra-low-noise timing jitter is usually obtained by locking its phase to a reference oscillator. 
High time-resolution measurement benefits from sampling one OFC signal with another OFC that has a slight repetition frequency offset ${\Delta f_\text{r}}$. The effective time-measurement result is theoretically optimized by a magnifying factor $\alpha = f_\text{r}/\Delta f_\text{r}$. Specifically, the envelope phase variation of each comb is enlarged and revealed on the macroscopic signal with the relation
\begin{equation}
	\Delta T = \alpha \Delta \tau,
\end{equation}
where $\Delta T$ is the phase-time fluctuation of the LOS signal, which effectively stretches the time interval $\Delta\tau$ of the under-test signal by $\alpha$ times in the time domain. \rev{It is worth mentioning that each LOS cycle consists of 100,000 pairs of comb pulses, and it requires reproducibility in both comb signals.}

Benefiting from the high-resolution LOS technique, here, we realize the fs-level-precision electrical time-interval measurement of two pulses with an arbitrary waveform by transferring the microwave signal from the electrical region to the optical region, and then the system enjoys the effective time stretching. For this purpose, both experimental and theoretical demonstrations showed that a simple structure with $f_\text{CEO}$-free was sufficient to achieve fs-level precision.

\textbf{Comb.} Self-developed SESAM-based OFCs are built with a 1,560-nm central wavelength and 12-nm line width. 
The repetition frequency $f_\text{r}$ is at approximately 100 MHz, which is adjusted by a piezoelectric ceramic transducer (PZT) with a bandwidth of 100 Hz, an optical delay line with an adjustable range up to 1 MHz and 100 Hz step, and a temperature-controlling plate with a 10 K temperature-controlling range and 0.001 K precision.

The OFCs have 100 mW of output power, which is achieved with a built-in Erbium-doped fiber amplifier (EDFA). The high output power ensures that the energy is sufficient for the subsequent operations, including phase locking, interference, and monitoring. 
The polarization of the OFCs is stabilized with the polarization controllers, for the stability of the interference.
The CEO frequency $f_\text{CEO}$ of the combs does not need to be controlled, so the system complexity can be significantly reduced.

\textbf{Signal preprocessing.}
To achieve the fs-precision time-interval measurement, we expect the phase of under-test electrical signals to be transferred to the corresponding OFCs precisely to employ the follow-up LOS technique, and then we can expect the interval of the combs to be stretched by LOS and precisely measured based on the combs' high time-resolution characteristics. The signal preprocessing procedure aims to filter \ydr{(together with multiplying/dividing)} a single frequency component with no phase distortion, which can serve as the reference oscillator of combs A and B.

Assume that two arbitrary signals have the forms given by 
\begin{align}
	{s_{{\rm{U}}1}} \left( t \right) &= \sum\limits_n {{g_1}\left( {t - n{t_{{\rm{U}}1}} - {\tau_1}} \right)} ,  \\
	{s_{{\rm{U2}}}} \left( t \right) &= \sum\limits_n {{g_2}\left( {t - n{t_{{\rm{U}}2}} - {\tau_2}} \right)} ,
\end{align}
where ${g_1}\left( t \right)$ and ${g_2}\left( t \right)$ are envelopes of two under-test signals, ${{t_{{\rm{U}}1}}}$ and ${{t_{{\rm{U}}2}}}$ denote periods of two signals, and $\tau_1$ and $\tau_2$ denote their phase times. The under-test time interval is the phase-time difference, i.e. $\Delta\tau = \tau_1-\tau_2$. 
\rev{When the under-test signals with fixed delays have the same frequency, one can obtain fixed values when measuring the phase time, phase, and delay between peaks. Moreover, it is important to note that when two signals with fixed delays have different frequencies, their phase time remains constant. However, a phase slip occurs when trying to determine the delay between two sets of peaks. To avoid ambiguity, it is more appropriate to use the phase time difference to represent the time interval of the periodic signals, rather than to use the delay between peaks (or equivalently, the phase $2\pi f\tau$) directly.}
After the signals pass through the BPFs with the transfer function of $h_i(t)$ and $h_j(t)$, we can obtain the quasi-single-frequency signals, given by
\begin{align}
	&s_1(t) = s_{\text{U}1}(t) * h_i(t) = \sin\left[2\pi if_{\text{U}1}(t-\tau_1)\right],\\
	&s_2(t) = s_{\text{U}2}(t) * h_j(t) = \sin\left[2\pi jf_{\text{U}2}(t-\tau_2)\right],
\end{align}
where $i,\ j$ are integers quantifying the $i$-\textit{th} (or $j$-\textit{th}) harmonics of the measured signals, respectively, and ${f_{{\rm{U}}1}} = {1 \mathord{\left/
		{\vphantom {1 {{t_{{\rm{U}}1}}}}} \right.
		\kern-\nulldelimiterspace} {{t_{{\rm{U}}1}}}}$ and ${f_{{\rm{U2}}}} = {1 \mathord{\left/
		{\vphantom {1 {{t_{{\rm{U2}}}}}}} \right.
		\kern-\nulldelimiterspace} {{t_{{\rm{U2}}}}}}$ are the frequencies of the under-test signals.
\ydr{To realize synchronization with the combs, the frequency of $if_\text{U1}$ and $jf_\text{U2}$ need to be manually filtered and multiplied to be in the bandwidth of the locking system (i.e. near integer multiples of 100 MHz).}

Note that for arbitrary waveform signals, different harmonics have different energies. 
\rev{For situations where direct filtering of the high-order harmonic is not feasible, an alternative approach is to obtain it by multiplying the first-harmonic signal, which typically possesses the highest energy.}
\rev{Furthermore, because the introduced time interval resulting from optical-microwave synchronization using FLOM-PD is unaffected by the power levels of the electrical signal, harmonic energy differences will not influence subsequent experiments. This characteristic enables our scheme to measure different waveform signals or use different frequency components without compromising the measurement results.} 
\\

\textbf{Conversion from electrical to optical signals.}
To fully utilize the high-time-resolution characteristic of OFCs, we must ensure that the OFCs carry the time information of the electrical under-test signals by phase-locking techniques. To realize high-precision phase locking, we use the FLOM-PDs, whose noise property surpasses the traditional mixer technique~\cite{RN892}.
The output of each OFC is split with a 50:50 beam splitter (BS), and the power of each arm is controllable with an optical attenuator. The branch used for phase-locking is set to 8 mW, ensuring sufficient energy of the error signal, and the one used for LOS interference is set to 20 mW so that the signal is large enough after the optical filter.

Using a self-developed repetition frequency controlling scheme, the locking system can operate continuously for over a month without interruption under the condition that the ambient temperature exceeds 5 K, making the system suitable for large temperature changes.

\textbf{Data acquisition and post-processing.} For data processing, the LOS signals' envelopes $A_i(t)$ ($i = 1, 2$ is the signal number) carry the time information of the under-test signals, so they are preliminarily extracted with the Hilbert transformation.
\begin{equation}
	\begin{aligned}
		A_i(t)=&\mathcal{H}[s(t)]=h(t)*s(t)\\
		=&\frac{1}{\pi}\int_{-\infty}^{\infty}\frac{s(\tau)}{t-\tau}\mathrm{d}\tau. 
	\end{aligned}
\end{equation}
Conventionally, the delay can be extracted with \ydr{curve-fitting} on the envelope and peak-finding, while the line shape of the LOS envelope depends on the OFCs' mode-locking process and needs optimization for different OFCs, such as hyperbolic secant fitting for an LOS signal generated by two soliton mode-locked lasers. If the gain narrowing effect of the amplifiers is considered, the line shape can be approximated to Gaussian~\cite{RN113}. In addition, the curve fitting is time-consuming, especially for a large data set. It is thus unfavorable for real-time processing of the data. 
Here, we use a more efficient and precise way, namely, calculating the envelopes' center of gravity with the definition
\begin{equation}
	t_i = \frac{\int_{-\infty}^\infty A_i(t')\times t'\mathrm{d}t'}{\int_{-\infty}^\infty A_i(t')\mathrm{d}t'},\  i = 1,2,
\end{equation}
where $t'$ represents the time axis, and the result is shown in FIG.~\ref{signalconverting} with black dots. 
Repeating the above process, one can continuously measure the time differences in \ydr{real-time}. 
To optimize the system precision, we designed a single-dimension Kalman filter to realize real-time processing of the time-interval data and restrain the white noise induced by the \rev{localization} process.

\rev{Additionally, a full scan of the LOS waveform covers multi-ns oscillator time and most of it measures baseline. To improve the efficiency of our scheme, we implemented a threshold-based approach. If a data point is determined to be a baseline, we do not record it but only increase the index of the data, thus optimizing the data processing speed in our implementation. Optimizing the measurement duty cycle would be another effective way, by using e.g., time-programmable frequency combs~\cite{dist1}. And we will leave this question for future studies.}

\textbf{System model.}
To give a further explanation of our system, one can consider the signal of comb B and comb X in a specific optical channel (about 0.8 nm in width). Both combs are composed of $N=1,000$ comb teeth, and the fluctuations of their repetition frequency and CEO frequency are taken into account. The signal of comb B can be written as
\small{
\begin{equation}
    \begin{aligned}
		&s_\text{B}(t-T) \\=& \sum_{n=0}^{N-1} \cos\left\{\int_0^{t-T}\left[(\omega_0^\text{B} +\delta\omega_0^\text{B}(t'))+n(\omega_\text{r}+\delta\omega_\text{r}^\text{B}(t'))\right]\mathrm{d}t'\right\},\\
		=& \sum_{n=0}^{N-1} \cos\left\{(\omega_0^\text{B}+n\omega_\text{r})(t-T)+\int_0^{t-T}[\delta\omega_0^\text{B}(t')+n\delta\omega_\text{r}^\text{B}(t')]\mathrm{d}t'\right\},
    \end{aligned}
    \label{combb}
\end{equation}
}
and similarly, that of the comb X can be written as
\small{
\begin{equation}
    \begin{aligned}
		&s_\text{X}(t)\\ =& \sum_{n=0}^{N-1} \cos\left\{[\omega_0^\text{X}+n(\omega_\text{r}+\Delta\omega_\text{r})]t+\int_0^{t}[\delta\omega_0^\text{X}(t')+n\delta\omega_\text{r}^\text{X}(t')]\mathrm{d}t'\right\},
    \end{aligned}
    \label{combx}
\end{equation}
}
where $\omega_0$ and $\delta\omega_0(t)$ are the lowest angular frequency in the spectra channel and its fluctuation, respectively, which are determined by $f_\text{CEO}$ and its fluctuation. $\omega_\text{r}$ and $\delta\omega_\text{r}(t)$ are the repetition angular frequency and its fluctuation, respectively. $\Delta\omega_\text{r}$ is the repetition frequency offset, and $N$ is the number of the comb teeth in the channel.
Interfering in the coupler, the signal detected by the BPD is the multiplication of $s_\text{B}(t-T)$ and $s_\text{X}(t)$, i.e.,
\small{
\begin{equation}
    \begin{aligned}
	&I(t)=s_\text{B}(t-T)\times s_\text{X}(t),\\
		=&\sum_{n\Delta\omega_\text{r}<2\pi B}\cos\left\{(\omega_0^\text{X}-\omega_0^\text{B}+n\Delta\omega_\text{r})\right.\\
    &\left.\left[ t+\frac{\omega_\text{r}}{\Delta\omega_\text{r}}\left(T+\frac{1}{\omega_\text{r}}\int_0^t\delta\omega_\text{r}^\text{X}(t')\mathrm{d}t'-\frac{1}{\omega_\text{r}}\int_0^{t-T}\delta\omega_\text{r}^\text{B}(t')\mathrm{d}t'\right)  \right]\right.\\
		&\left.+\int_0^t\delta\omega_0^\text{X}(t')\mathrm{d}t'-\int_0^{t-T}\delta\omega_0^\text{B}(t')+(\omega_0^\text{X}-\omega_0^\text{B})\mathrm{d}t'\right\},\\
		=& \sum_{n\Delta\omega_\text{r}<2\pi B}\cos\left\{(\Delta\omega_0+n\Delta\omega_\text{r})\right.\\
        &\left.\left[ t+\frac{\omega_\text{r}}{\Delta\omega_\text{r}}\left(T+x^\text{X}(t)-x^\text{B}(t-T)\right)\right]+\varphi_0^\text{X}(t)-\varphi_0^\text{B}(t-T)\right\},
    \end{aligned}
    \label{loseq}
\end{equation}
}
where $x(t)=\frac{1}{\omega_\text{r}}\int_0^t\delta\omega_\text{r}(t')\mathrm{d}t'$ is the phase time fluctuation induced by the repetition frequency fluctuation, $\varphi_0(t)=\int_0^t\delta\omega_0(t')\mathrm{d}t'$ indicates the phase noise induced by the frequency fluctuation of $\omega_0$, and $B$ is the bandwidth of BPD. 
From this result, one can see that the LOS signal has a similar form as a comb signal, and its repetition frequency is the repetition frequency offset of the two combs $\Delta\omega_\text{r}$. Moreover, the delay of the signal under test $T$ and the phase time fluctuation induced by $\delta\omega_\text{r}(t)$ of both of the combs are enlarged by a factor of $(\omega_\text{r}/\Delta\omega_\text{r})$.
Considering the noise property, we determine the frequency stability of the LOS signal by both of the combs' repetition frequency stability.
Statistically and precisely, because the two noise sources are independent with each other, one can write 
\begin{equation}
	\langle x^\text{LOS}(t)\rangle=\frac{\omega_\text{r}}{\Delta\omega_\text{r}}\sqrt{\langle x^\text{X}(t)\rangle^2+\langle x^\text{B}(t)\rangle^2}.
	\label{PNrelation}
\end{equation}
This means that the measured single LOS signal frequency stability is determined by the relative repetition frequency fluctuation of comb B, comb X, and the enlargement factor $(\omega_\text{r}/\Delta\omega_\text{r})$.

Together with the LOS signal of comb X and A, the relative fluctuation of their phase time difference, that is, the system limit, can be written as 
\begin{equation}
	\langle x^T(t)\rangle=\frac{\omega_\text{r}}{\Delta\omega_\text{r}}\sqrt{\langle x^\text{A}(t)\rangle^2+\langle x^\text{B}(t)\rangle^2},
\end{equation}
which indicates that the phase time fluctuation of comb X can be omitted on the condition that the difference in the delay from comb X to comb A and B can be omitted, which can be easily fulfilled.

In addition, from Eq.~\eqref{combb}, Eq.~\eqref{combx} and Eq.~\eqref{loseq}, it is seen that the form of an LOS signal is exactly the same as that of a comb signal. Furthermore, $\Delta\omega_0$ of the LOS signal corresponds to $\omega_0$ of a comb signal, which indicates that $\Delta\omega_0$ is the carrier frequency of the LOS signal and does not influence the envelopes' phase time.
By contrast, intuitively, to ensure the \rev{localization} precision, the system ensures that $\Delta\omega_0$ is intimately linked to the sampling rate. Based on the simple formula of Eq.~\eqref{loseq}, we simulated the influence of $\Delta\omega_0$, the sampling rate, and RIN on the \rev{localization} precision. The results are shown in FIG.~\ref{3d}.

Over and above that, Eq.~\eqref{combb} is also a powerful tool to describe not only its own comb signal but also those affected by fiber dispersion and other nonlinear effects.


\section{backmatter}

\textbf{Funding.}
This work was supported by the National Natural Science Foundation of China (Grants Nos. 62201012, and 61531003), and the China Postdoctoral Science Foundation (Grant No. 2020TQ0016).

\textbf{Author contributions.}
All authors contributed to the scientific discussions of the study. Z.C., B.L., and H.G. conceived the study. D.Y. and Z.C. developed the experimental plan, D.Y., Z.C., Y.X., Z.J., P.M. and Y.Z. designed, built, and experimentally validated the time-interval measurement setup and performed the data analysis. S.Y., B.L., and H.G. discussed the optimization of the experiment scheme. D.Y. and X.Y. perform the data processing. D.Y. carried out the theoretical calculations and the simulation. D.Y. and Z.C. co-wrote the manuscript, and all authors provided revisions.

\textbf{Disclosures.}
The authors declare no conflicts of interest.

\textbf{Data availability.} Data underlying the results presented in this paper are not publicly available at this time but may be obtained from the authors upon reasonable request.




\begin{thebibliography}{99}  
	

	\bibitem{timesync1} Q. Shen, J.-Y. Guan, J.-G. Ren, T. Zeng, L. Hou, M. Li, Y. Cao, J.-J. Han, M.-Z. Lian, Y.-W. Chen, X.-X. Peng, S.-M. Wang, D.-Y. Zhu, X.-P. Shi, Z.-G. Wang, Y. Li, W.-Y. Liu, G.-S. Pan, Y. Wang, Z.-H. Li, J.-C. Wu, Y.-Y. Zhang, F.-X. Chen, C.-Y. Lu, S.-K. Liao, J. Yin, J.-J. Jia, C.-Z. Peng, H.-F. Jiang, Q. Zhang, and J.-W. Pan, “Free-space dissemination of time and frequency with $10^{-19}$ instability over 113 km,” Nature \textbf{610}, 661–666 (2022).

	\bibitem{timesync2} J.-D. Deschênes, L. C. Sinclair, F. R. Giorgetta, W. C. Swann, E. Bau- mann, H. Bergeron, M. Cermak, I. Coddington, and N. R. Newbury, “Synchronization of distant optical clocks at the femtosecond level,” Phys. Rev. X \textbf{6}, 021016 (2016).

	\bibitem{timesync3} M. Xin, K. Safak, and F. X. Kärtner, “Ultra-precise timing and synchro- nization for large-scale scientific instruments,” Optica \textbf{5}, 1564–1578 (2018).

	\bibitem{timesync4} Q. Shen, J.-Y. Guan, T. Zeng, Q.-M. Lu, L. Huang, Y. Cao, J.-P. Chen, T.-Q. Tao, J.-C. Wu, L. Hou, S.-K. Liao, J.-G. Ren, J. Yin, J.-J. Jia, H.-F. Jiang, C.-Z. Peng, Q. Zhang, and J.-W. Pan, “Experimental simulation of time and frequency transfer via an optical satellite-ground link at $10^{-19}$ instability,” Optica \textbf{8}, 471–476 (2021).

	\bibitem{navig1} Y. Shen, S. Mazuelas, and M. Z. Win, “Network navigation: theory and interpretation,” IEEE J. Sel. Areas Commun. \textbf{30}, 1823–1834 (2012).

	\bibitem{navig2} J. Neil, L. Cosart, and G. Zampetti, “Precise timing for vehicle navigation in the smart city: an overview,” IEEE Commun. Mag. \textbf{58}, 54–59 (2020).

	\bibitem{navig3} B. Jaduszliwer and J. Camparo, “Past, present and future of atomic clocks for GNSS,” GPS Solutions \textbf{25}, 27 (2021).

	\bibitem{navig4} P. Zhang, R. Tu, X. Lu, Y. Gao, and F. Lihong, “Performance of global positioning system precise time and frequency transfer with integer ambiguity resolution,” Meas. Sci. Technol. \textbf{33}, 045005 (2022).

	\bibitem{highenergy1} Q. An, “Review of methods and techniques of precise time interval measurements for particle physics experiments,” Nucl. Tech. \textbf{29}, 453–462 (2006).

	\bibitem{darkmatter1} B. M. Roberts, G. Blewitt, C. Dailey, M. Murphy, M. Pospelov, A. Rollings, J. Sherman, W. Williams, and A. Derevianko, “Search for domain wall dark matter with atomic clocks on board global positioning system satellites,” Nat. Commun. \textbf{8}, 1195 (2017).

	\bibitem{darkmatter2} N. Huntemann, B. Lipphardt, C. Tamm, V. Gerginov, S. Weyers, and E. Peik, “Improved limit on a temporal variation of $m_p/m_e$ from comparisons of Yb$^+$ and Cs atomic clocks,” Phys. Rev. Lett. \textbf{113}, 210802 (2014).

	\bibitem{darkmatter3} A. Derevianko and M. Pospelov, “Hunting for topological dark matter with atomic clocks,” Nat. Phys. \textbf{10}, 933–936 (2014).

	\bibitem{darkmatter4} J. Liu, X. Chen, and X. Ji, “Current status of direct dark matter detection experiments,” Nat. Phys. \textbf{13}, 212–216 (2017).

	\bibitem{const1} J. C. Berengut and V. V. Flambaum, “Testing time-variation of fundamental constants using a $^{229}$Th nuclear clock,” Nuclear Phys. News \textbf{20}, 19–22 (2010).

	\bibitem{const3} M. Safronova, D. Budker, D. DeMille, D. F. J. Kimball, A. Derevianko, and C. W. Clark, “Search for new physics with atoms and molecules,” Rev. Mod. Phys. \textbf{90}, 025008 (2018).

	\bibitem{const4} Y. V. Stadnik and V. V. Flambaum, “Searching for dark matter and variation of fundamental constants with laser and maser interferometry,” Phys. Rev. Lett. \textbf{114}, 161301 (2015).

	\bibitem{RN2390} K. Józef, “Review of methods for time interval measurements with picosecond resolution,” Metrologia \textbf{41}, 17 (2004).

	\bibitem{RN2377} I. P. Dan, “Review of sub-nanosecond time-interval measurements,” IEEE Trans. Nucl. Sci. \textbf{20}, 36–51 (1973).

	\bibitem{RN2381} J. Zhao, Z. Zhao, and L. Fu, “Research on the high resolution precision time-interval measurement methods,” Proc. Eng. \textbf{174}, 1257–1261 (2017).

	\bibitem{RN2392} S. Henzler, \textit{Time-to-Digital Converter Basics} (Springer, 2010).

	\bibitem{RN2386} X. Ren and X. F. Zhang, “Methods of high precision time-interval measurement,” in \textit{4th International Conference on Electronic Information Technology and Computer Engineering (EITCE) }(2020).

	\bibitem{RN2391} J. P. Jansson, A. Mantyniemi, and J. Kostamovaara, “A CMOS time-to-digital converter with better than 10 ps single-shot precision,” IEEE J. Solid-State Circuits \textbf{41}, 1286–1296 (2006).

	\bibitem{RN1156} D. W. Allan and H. Daams, “Picosecond time difference measurement system,” in \textit{Symposium on Frequency Control} (1975).

	\bibitem{OFC1} N. R. Newbury, “Searching for applications with a fine-tooth comb,” Nat. Photonics \textbf{5}, 186–188 (2011).

	\bibitem{OFC2} J. L. Hall, “Nobel lecture: defining and measuring optical frequencies,” Rev. Mod. Phys. \textbf{78}, 1279–1295 (2006).

	\bibitem{RN2266} H. Margolis, G. Barwood, G. Huang, H. Klein, S. Lea, K. Szymaniec, and P. Gill, “Hertz-level measurement of the optical clock frequency in a single $^{88}$Sr$^+$ ion,” Science \textbf{306}, 1355–1358 (2004).

	\bibitem{RN2265} S. A. Diddams, T. Udem, J. Bergquist, E. Curtis, R. Drullinger, L. Hollberg, W. M. Itano, W. Lee, C. Oates, and K. Vogel, “An optical clock based on a single trapped $^{199}$Hg$^+$ ion,” Science \textbf{293}, 825–828 (2001).

	\bibitem{LOSsync1} H. Bergeron, L. C. Sinclair, W. C. Swann, I. Khader, K. C. Cossel, M. Cermak, J.-D. Deschênes, and N. R. Newbury, “Femtosecond time synchronization of optical clocks off of a flying quadcopter,” Nat. Commun. \textbf{10}, 1819 (2019).

	\bibitem{LOSsync2} Q. Lu, Q. Shen, J. Guan, M. Li, J. Chen, S. Liao, Q. Zhang, and C. Peng, “Sensitive linear optical sampling system with femtosecond precision,” Rev. Sci. Instrum. \textbf{91}, 035113 (2020).

	\bibitem{dist1} E. D. Caldwell, L. C. Sinclair, N. R. Newbury, and J.-D. Deschenes, “The time-programmable frequency comb and its use in quantum-limited ranging,” Nature \textbf{610}, 667–673 (2022).

	\bibitem{dist2} M. Kajima and K. Minoshima, “A simple optical-zooming positioning system using a femtosecond frequency comb,” in \textit{Conference on Lasers \& Electro-Optics} (2009).

	\bibitem{dist3} M. Kajima and K. Minoshima, “Optical zooming interferometer for subnanometer positioning using an optical frequency comb,” Appl. Opt. \textbf{49}, 5844–5850 (2010).

	\bibitem{dist4} I. Coddington, W. C. Swann, L. Nenadovic, and N. R. Newbury, “Rapid and precise absolute distance measurements at long range,” Nat. Photonics \textbf{3}, 351–356 (2009).

	\bibitem{RN928} G. Marra, “Transfer of optical frequency combs over optical fibre links,” Thesis (University of Southampton, 2013).

	\bibitem{RN964} W. Zhang, T. Li, M. Lours, S. Seidelin, G. Santarelli, and Y. L. Coq, “Amplitude to phase conversion of InGaAs pin photo-diodes for femtosecond lasers microwave signal generation,” Appl. Phys. B \textbf{106}, 301–308 (2012).

	\bibitem{RN2539} Z. Jin, Y. Xu, D. Yu, B. Luo, Z. Chen, G. Wu, and H. Guo, “Analyzing the influence of InGaAs photodetectors in comb-based frequency transfer,” in \textit{Frontiers in Optics + Laser Science} (2022).

	\bibitem{FLOM-PD1} J. Kim, F. X. Kärtner, and M. H. Perrott, “Femtosecond synchronization of radio frequency signals with optical pulse trains,” Opt. Lett. \textbf{29}, 2076–2078 (2004).

	\bibitem{RN892} K. Jung and J. Kim, “Subfemtosecond synchronization of microwave oscillators with mode-locked Er-fiber lasers,” Opt. Lett. \textbf{37}, 2958–2960 (2012).

	\bibitem{RN1150} J. Kim, J. A. Cox, J. Chen, and F. X. Kärtner, “Drift-free femtosecond timing synchronization of remote optical and microwave sources,” Nat. Photonics 2, 733–736 (2008).

	\bibitem{RN2333} F. R. Giorgetta, W. C. Swann, L. C. Sinclair, E. Baumann, I. Coddington, and N. R. Newbury, “Optical two-way time and frequency transfer over free space,” Nat. Photonics \textbf{7}, 434–438 (2013).

	\bibitem{RN113} G. P. Agrawal, \textit{Nonlinear Fiber Optics} (Springer, 2000).





\end{thebibliography}
\end{document}